# Modeling community standards for metadata as templates makes data FAIR

Mark A. Musen[1], Martin J. O'Connor[1], Erik Schultes[2], Marcos Martínez-Romero[1*], Josef Hardi[1], and John Graybeal[1]

[1]Stanford Center for Biomedical Informatics Research
Stanford University
Stanford, CA 94305 USA

[2]GO FAIR Foundation
Rijnsburgerweg 10
2333 AA Leiden NL

**Abstract**
It is challenging to determine whether datasets are findable, accessible, interoperable, and reusable (FAIR) because the FAIR Guiding Principles refer to highly idiosyncratic criteria regarding the metadata used to annotate datasets.  Specifically, the FAIR principles require metadata to be "rich" and to adhere to "domain-relevant" community standards.  Scientific communities should be able to define their own *machine-actionable templates* for metadata that encode these "rich," discipline-specific elements.  We have explored this template-based approach in the context of two software systems.  One system is the CEDAR Workbench, which investigators use to author new metadata.  The other is the FAIRware Workbench, which evaluates the metadata of archived datasets for their adherence to community standards.  Benefits accrue when templates for metadata become central elements in an ecosystem of tools to manage online datasets—both because the templates serve as a community reference for what constitutes FAIR data, and because they embody that perspective in a form that can be distributed among a variety of software applications to assist with data stewardship and data sharing.

**Introduction**
In 2014, a small workshop of invited participants seized on the growing recognition in the scientific community that research data should be made available in open repositories and, furthermore, that the data should be archived in a manner that makes them maximally usable by other investigators.  The data needed to be FAIR—findable, accessible, interoperable, and reusable by and for machines as well as people—and, after the publication of that fitting acronym a short time later [1], funders, publishers, professional groups, and investigators

---

[*]Current Address:  Acubed Innovation Center, 601 West California Avenue, Sunnyvale, CA  94086 USA

# The FAIR Guiding Principles

| | |
|---|---|
| **F1:** **(Meta) data** are assigned globally unique and persistent identifiers | **I1:** **(Meta)data** use a formal, accessible, shared, and broadly applicable language for knowledge representation |
| **F2:** Data are described with <u>rich **metadata**</u> | **I2:** **(Meta)data** use vocabularies <u>that follow the FAIR principles</u> |
| **F3:** **Metadata** clearly and explicitly include the identifier of the data they describe | **I3:** **(Meta)data** include qualified references to other (meta)data |
| **F4:** **(Meta)data** are registered or indexed in a searchable resource | |
| **A1:** **(Meta)data** are retrievable by their identifier using a standardized communication protocol | **R1:** **(Meta)data** are <u>richly described with a plurality of accurate and relevant attributes</u> |
| **A1.1:** The protocol is open, free, and universally implementable | **R1.1:** **(Meta)data** are released with a clear and accessible data usage license |
| **A1.2:** The protocol allows for an authentication and authorization where necessary | **R1.2:** **(Meta)data** are associated with <u>detailed provenance</u> |
| **A2:** **Metadata** should be accessible even when the data is no longer available | **R1.3:** **(Meta)data** meet <u>domain-relevant community standards</u> |

**Table 1:** The FAIR Guiding Principles as presented by Wilkinson et al. [1]. Most of the principles relate to metadata. In the Table, underlining indicates subjective aspects of the FAIR principles that are community dependent.

themselves all began to see the ability to disseminate, to access, and to analyze FAIR research data as an important goal. Suddenly, funders adopted policies that their grantees' data must be FAIR as well as archived in open repositories [2]. Some journals would not publish investigators' papers (or even referee them) unless the data were FAIR [3]. Not surprisingly, the trend over the past few years has been for many researchers simply to assert that their data are FAIR, when in fact they often fail to adhere to the FAIR principles. Such claims of "FAIRness" have led to inflation of the concept, and they risk its ultimate devaluation at a time when FAIR data and data sharing are increasingly needed.

Operationalizing the FAIR Guiding Principles
Even though the terms *findable*, *accessible*, *interoperable*, and *reusable* have clear, vernacular meaning, the "FAIR Guiding Principles"—which were published in the same article that called on the scientific community to ensure that data are FAIR [1]—are rather loosely defined (Table 1). Many of the FAIR Guiding Principles relate to issues that typically are handled by the *repositories* where the data are stored (e.g., the use of globally unique and persistent identifiers, the ability to search the repository, the use of standard communication protocols). Such matters are therefore out of the hands of the investigators who are asked to place their data in a particular archive.

The FAIR principles over which investigators *do* have control deal with the metadata. Are the data described with rich metadata? Are the metadata richly described with a plurality of accurate and relevant attributes? Do the metadata meet domain-relevant community standards? It is therefore important that investigators pay particular attention to such considerations if they want their datasets to be FAIR. The challenge, however, is that most investigators and data curators don't know how to begin to operationalize such principles in the setting of annotating particular datasets. What exactly makes metadata attributes "accurate and relevant"? What qualifies a metadata specification as "rich"? Which community standards are the most important? The challenge with the FAIR Guiding Principles is that they are abstract and lack context. Although their high level of abstraction may have contributed to their rapid endorsement by the scientific community, it has been very hard for investigators to operationalize the FAIR principles, and for third parties—in particular, policy makers—to determine when the principles have been implemented successfully.

Faced with this confusion, investigators have tried to evaluate whether datasets adhere to the FAIR principles by using automated tools that examine the data records in particular repositories and grade them for "FAIRness." Computer programs such as the FAIR Evaluator [4], FAIRShake [5], F-UJI [6], and many others have attempted to automate the task of determining whether data records are FAIR. Despite the intense interest in making data FAIR, these systems have yet to accrue large followings because it is impossible for such programs to determine autonomously whether many of the FAIR principles are even being followed. How does a computer know whether datasets are being annotated with metadata that are sufficiently "rich"? How does a computer know which community standards for metadata are germane and whether they have been applied correctly? Entwined with the idea of FAIR data is the human-centered problem of ensuring that metadata include sufficient descriptors so that members of a particular scientific community can find the datasets that they are looking for with reasonable recall and precision, and that those metadata descriptors include the kinds of standard terms that different groups of investigators will employ in their queries.

If the goal is to provide automated support for creating FAIR data and for evaluating existing datasets for FAIRness, then we need a mechanism to define for the computer all the attendant human-centered issues. We need a way to specify which communities and which community-based standards matter in a particular context, and how metadata authors are to be expected to follow them. We need a way to specify which standard vocabularies are preferred for supplying the terms used in our metadata. We need a way to specify what constitutes a "rich" set of attributes and what represents a "rich" set of descriptions for a given attribute. Without explicit definitional knowledge, none of these notions is inferable automatically. In the work that we present in this paper, we propose that communities of investigators should create machine-processable metadata *templates* that embody their own relevant standards and that guide data stewards in how those standards should be applied.

Our metadata templates comprise descriptions of the attribute–value pairs that characterize standard metadata specifications. The templates are represented in a standard machine-readable language (JSON Schema) and they encode the community-based standards typically needed to create research metadata in a consistent manner. The metadata templates therefore capture for the computer all the subjective elements needed to operationalize the FAIR principles in particular domains of science, for particular types of experiments, and for particular communities of experimenters (Figure 1).

**Figure 1:** Metadata template for capturing information about a tissue sample. This screen capture shows the template used by investigators in the NIH-supported HuBMAP consortium to specify metadata about biological specimens used to perform assays of cell-specific biomarkers. In the figure, the user is entering a controlled term from a special HuBMAP ontology to provide the metadata entry for the specimen's preparation medium. The attributes of tissues are the ones that the HuBMAP community has chosen to standardize for its descriptions of such samples. The ontology terms used to provide values for the metadata attributes similarly represent community-endorsed standards for declaring this kind of information.

A metadata template allows an investigator to describe all the "data about the data" needed to understand the nature of a study, its motivation, and the means by which the study was executed. Such a template encapsulates in a single, machine-readable place everything that a third party—or a computer—needs in order to interpret what has been done and whether the data are reusable in a given context. A filled-out metadata template is analogous to an electronic cartridge that a user might plug into a gaming console or into a music synthesizer to transmit formal, standardized information quickly and completely—altering the behavior of the system receiving the information so that the system reflects the information that the cartridge is designed to communicate. People use electronic cartridges and similar devices all the time to bundle information and to permit information reuse and dissemination—particularly when the information is extensive, complex, or hard to articulate. We see the same need for encapsulating community-dependent, context-specific information about scientific experiments in order to guarantee that the metadata are "rich" and complete and that the resulting datasets are FAIR.

Metadata specifications are not physical artifacts like a cartridge or a cassette, of course, but they are discrete, self-contained, digital research objects. Whenever multiple computer applications are needed in a data ecosystem (e.g., to create new metadata descriptions that ensure the FAIRness of the underlying datasets, to assess how well existing datasets adhere to the FAIR principles, to offer recommendations regarding data repositories that are hoping to archive FAIR data), users should be able simply to "plug in" new cartridges to transmit directly to the computer the particular community-based FAIR criteria against which the users would like the computer to perform its analysis.

In this paper, we describe the structure of our metadata templates, and we present how these templated representations of standard metadata frameworks drive two different applications: (1) the CEDAR Workbench (or simply CEDAR) [7], which helps investigators to author the rich, standards-adherent metadata needed to make datasets FAIR, and (2) the FAIRware Workbench (or simply FAIRware), which uses metadata templates to assess the degree to which existing online data resources actually are FAIR. The approach demonstrates the value of representing data standards in a declarative, machine-readable manner, and of building applications that can share these representations so that community-dependent knowledge of the standards and practices regarding their use can be transmitted seamlessly among the various tools.

When we speak of metadata in this paper, we typically are referring to data about experimental data. Metadata may also refer to descriptions of other types of digital objects, such as software packages and workflows. Although our emphasis in this paper is on making experimental datasets FAIR, the approach is quite general, and it applies to all types of metadata.

Standards for Encoding Metadata
Community standards for describing the data collected in the course of scientific experiments have been evolving since the end of the last century, and they are continuously expanding in number. Work in the area of clinical research provides a good example. Practically since the advent of controlled clinical trials, biostatisticians have lamented the frequent lack of methodological information in reports of medical experiments to offer readers detailed understanding of the experiment and to enable confidence in the results [8]. Intense discussions during the 1990s ultimately led to the CONsolidated Standards Of Reporting Trials (CONSORT), a 25-item checklist of the "minimal information" needed to make sense of data from randomized clinical trials and a flow diagram for applying the elements of the checklist [9]. The CONSORT criteria provided guidance to journal editors and reviewers who sought to know whether the reported data and methods of a clinical study were complete and consistent with the intended study design, the actual execution of the study, and the analysis of the data. The CONSORT checklist also informed the development of ClinicalTrials.gov, the primary repository for preregistration of intervention trials in the United States [10]. The information about individual studies in ClinicalTrials.gov can be viewed as *metadata* for those studies, as the entries for each trial provide the information needed to make sense of the data that ultimately are collected [11]. Many of the fields in ClinicalTrials.gov records correspond directly to items in CONSORT.

After the popularization of the CONSORT minimal-information checklist, the EQUATOR Network [8] built upon the CONSORT approach to offer the medical community more than 500 such *reporting guidelines* for enumerating the minimal information that should be provided with clinical studies, economic evaluations, meta-analyses, and other kinds of scholarly reports, which now are used routinely to ensure that authors record the kinds of things about their work that will enable third parties to make sense of the data and the way in which the authors have performed their analyses.

Minimal information checklists for interpreting experimental results have been increasingly important in other areas of science. For example, around the time that CONSORT was being described for clinical trials, the community of scientists studying functional genomics—exploring how gene function gets turned on and turned off in different situations—recognized that the then-emerging technology of DNA microarrays required the readers of journal articles to consider large amounts of information to make sense of the complex experiments that were being reported. A group of investigators proposed their own reporting guideline—the Minimal Information About a Microarray Experiment (MIAME)—that offered a checklist to make sure that a publication contained sufficient detail for a third party to make sense of the study and to understand the results [12]. MIAME reminds investigators that a "minimal" report needs to include facts about how the experimental results are encoded, the subject of the study, the experimental set-up, and so on. MIAME became important not only to journal editors who wanted to ensure the completeness of scientific articles, but also to the developers of scientific databases such as the Gene Expression Omnibus [13], where the elements of the MIAME checklist have been used as fields in the metadata that describe the associated microarray datasets [14]. Thus, as with clinical trials, the standard reporting guideline for the microarray community both supported the editorial management of journal submissions and served as the basis for the metadata in the definitive repository used in the discipline.

In the biology community, the notion of minimal information checklists caught on rapidly, with researchers working at the grass roots proposing scores of reporting guidelines, such as Minimal Information About T-cell Assays (MIATA) [15] and Minimal Information For In Situ Hybridization and Immunohistochemistry Experiments (MISFISHIE) [16]. Investigators in other areas of science, recognizing the direct relationship between rich metadata and the FAIR principles, proposed corresponding reporting guidelines for use in earth science [17], ecology [18], and other disciplines. The FAIRsharing resource maintained at Oxford University provides information on more than 200 reporting guidelines from many disciplines within the natural sciences, generally in the form of minimal information checklists [19].

Recently, computer scientists in the machine learning community have recognized that it is impossible to process datasets without minimal information about the data content. Workers at Microsoft Research have proposed the idea that datasets should be accompanied by a "datasheet" [20] that enumerates essential information about the data—metadata that have aspects similar to those of a minimal information checklist.

Traditional minimal information checklists cannot be processed by computers in a meaningful way. Most checklists exist as textual documents, written with all the ambiguity and imprecision of natural language [21]. These checklists enumerate the things that need to be said about experiments and their data, but they are not in a form that enables a computer to verify that a checklist has been followed; they are designed primarily so that humans can attempt to tick boxes to indicate that *something* has been mentioned about the given topic. In our work, we make it possible for scientific communities to recast minimal information checklists as machine-readable templates that enumerate the *attributes* about a type of experiment and its associated data and the types of *values* that those metadata components can take on—helping the investigator to comply with reporting guidelines while at the same time encoding this information in an explicit and unambiguous format. For example, Figure 1 shows a metadata template that defines a reporting guideline for information about a tissue specimen. The template was created by scientists working in the Human Bio-Molecular Atlas Program of the U.S. National Institutes of Health (HuBMAP) [22]. The template corresponds to a guideline that lists the minimal set of attributes that need to be reported for a tissue specimen obtained during a surgical procedure or at autopsy. The Figure shows the use of the template within the CEDAR workbench [7], which includes a tool that enables investigators to create instances of metadata (i.e., metadata for a particular experiment, which, in this case, annotate data about a specimen). CEDAR guides the user in filling in the fields of the template with values that adhere to predefined types. Some of these value types might be integers or dates. Some might be arbitrary text strings. In many cases, the values are terms that come from standard ontologies.

The ontologies used to standardize metadata entries are collections of controlled terms that represent the entities in a domain, along with information about the relationships among those entities [23]. In the sciences, standard ontologies such as the Gene Ontology [24] are often used by data curators to provide consistent vocabularies for data annotation. In the metadata templates that form the focus of our work, curators use reporting guidelines (such as the one in Figure 1) as an overall framework and they select specific terms from appropriate ontologies to ensure that the elements of the guideline are filled out in a consistent way. Whenever an element of a reporting guideline needs to take on a standardized value, the field in the corresponding metadata template designates one or more ontologies (or branches of these ontologies, or set of individual terms, or a combination of these options) that should be used to supply the values (see the drop-down menu in Figure 1) [25].

Addressing Mandates for Data FAIRess
Our work is taking place within the context of growing international sentiment that research data need to be FAIR. The European Union's most recent program for funding science, Horizon Europe, requires all data generated by its grantees to be FAIR. In the United States, the Office of Science and Technology Policy recently provided guidance to all U.S. funding agencies to have policies in place by 2025 to ensure equitable, freely available access to all research results and data. National funding agencies increasing view the results of the research that they support— including the data—as a public good, and they view the availability of FAIR data as the means to deliver to tax payers the benefit that they have paid for [26]. At the same time, scientific communities are seeking ways to make the results of their research more valuable to other investigators and to comply with the new, escalating requirements for FAIR data.

In this paper, we present the notion of *metadata templates* as a fundamental mechanism by which groups of investigators to capture and communicate their requirements for the metadata needed to make their datasets FAIR; to enable computers to assist them in authoring, evaluating, and publishing experimental metadata; and to ensure that their judgments and preferences regarding guidelines for reporting experimental results can be propagated throughout a ecosystem of tools and processes that, together, help to guarantee that their data are FAIR.

**Results: Putting Metadata Templates to Use**
Our approach renders reporting guidelines as machine-readable templates, and it designates the controlled terms that can supply values for the fields in the reporting guidelines as elements of standard ontologies. When scientific metadata are encoded in this manner, the result is a consistent, reusable "cartridge" that can be reapplied in different applications. Our laboratory has developed two such applications: CEDAR and FAIRware. We now discuss these two applications in turn.

Authoring Metadata with Templates: The CEDAR Workbench
The CEDAR Workbench is a Web-based platform that enables users to manage a library of metadata templates, to share templates with one another and with designated groups, and to fill out those templates to create instances of experiment-specific metadata. The system includes a component that enables developers to create and edit new metadata templates. Users construct such templates by piecing together descriptions of each of the template's fields, possibly adapting previously defined templates or elements of such templates. For each field, users specify the *value type* (e.g., integer, date, character string) and, when appropriate, they link string-valued fields to the ontologies (or portions of ontologies) archived in the BioPortal open ontology repository [27] that can designate the controlled terms that should be used when curators later author instances of metadata. Template developers can also link metadata fields to *common data elements* (CDEs) that encapsulate defined value types (including enumerated lists of elements from ontologies) with the standard questions to be posed to end users to acquire those values [28].

Because CEDAR provides an editor for creating these machine-actionable metadata templates, we typically refer to them as "CEDAR templates." However, an important driver of our work is that templates are independent of any software system. In fact, the templates can be created by using simple text editors or by using more advanced applications developed by other groups [21].

When a user wishes to create metadata to annotate a specific dataset, he or she selects the appropriate community-inspired template from the CEDAR library of metadata templates. Users obviously need to know which template to choose, and they use the name of the template to guide the selection (Figure 2). CEDAR then employs the indicated template to generate automatically a simple Web form that allows the user to fill in the values of the fields in the designated template one by one. The CEDAR Workbench ensures that the user enters values that match the indicated value type for the field. If the field is linked to an ontology (or set of ontologies), then CEDAR provides a drop–down menu of appropriate controlled terms from which the user may make a selection (see Figure 1). CEDAR maintains a cache of metadata that have been previously entered into the system and uses that cache to learn patterns in the metadata [29]. These patterns enable CEDAR to tailor its user interface to speed the entry of metadata information. For example, the drop-down menu in Figure 1 sorts the entries from which the user is to make a selection so that, in the context of previously entered metadata field information, the most likely selections appear at the top of the menu.

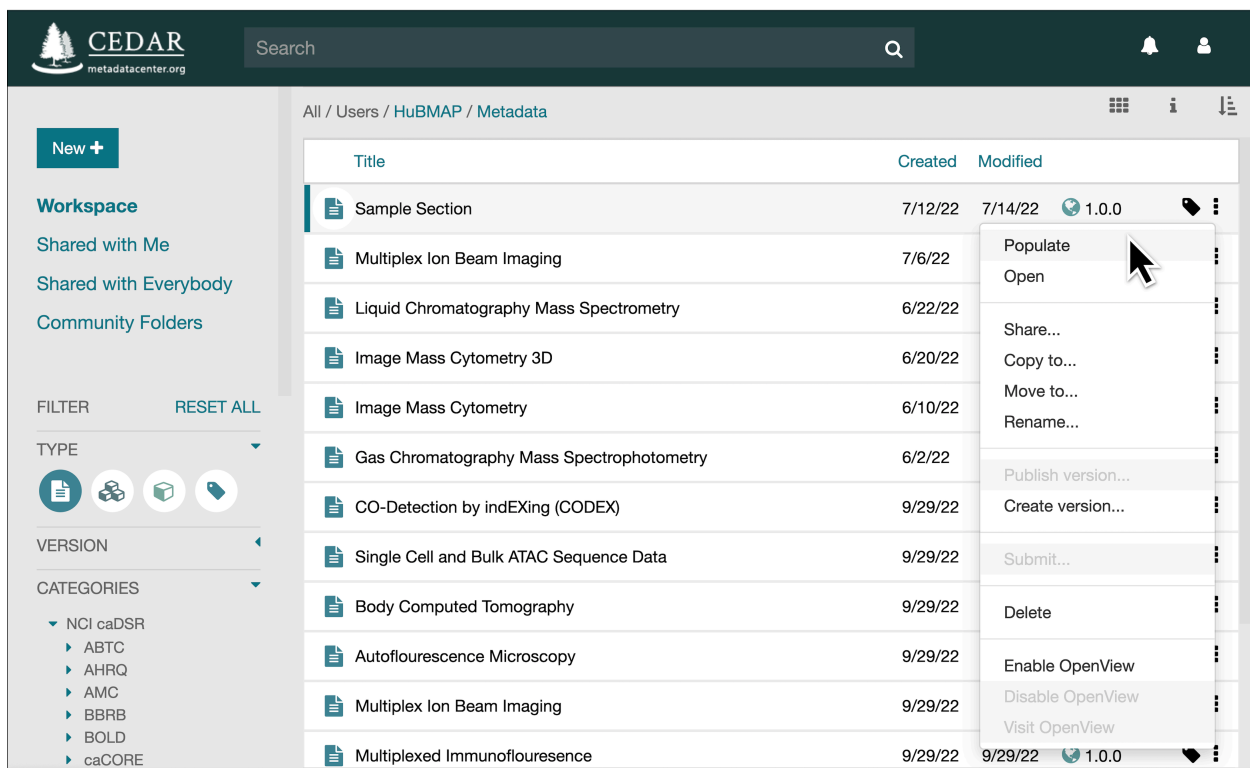

**Figure 2:** A collection of metadata templates in the CEDAR library. The screen capture depicts a set of templates created by HuBMAP users or shared with their community members. In CEDAR, users may view and access their own metadata templates, templates explicitly shared with the user by others, and templates shared by designated research communities stored in "community folders." Here, the user is seeking to populate the Sample Section template, which appears in Figure 1.

Creating metadata in CEDAR is a matter of filling out templates that reflect community-specific reporting guidelines. If the template is created with appropriate care, then the template adheres to community standards because the community was responsible for creating (or at least vetting) the template in the first place; the metadata will be as "rich" as the reporting guideline indicates they need to be, and they will use standard scientific ontologies that follow the FAIR principles. The subjective elements of the FAIR principles (i.e., adherence to appropriate standards, "richness") are built into the template specification. Users who complete the template will therefore have guarantees that their metadata instances are in full compliance with the community-determined standards. Thus, when CEDAR users fill in such a template and place the metadata and the corresponding data into a repository, the dataset will be, by definition, FAIR.

CEDAR can upload datasets to designated repositories that have application-program interfaces that have been mapped to the system. Depending on the setting, the dataset may be transmitted to a local data-coordinating center [30], to a generalist data repository, or to a domain-specific repository, such as those at the National Center for Biotechnology Information [31]. In the latter case, where the domain-specific repository may incorporate assumptions about the semantics of uploaded metadata, it is essential that the original CEDAR template be constructed to match those semantics.

Investigators currently are using CEDAR to author metadata in large NIH projects such as HuBMAP [22], the Library of Integrated Network-based Cellular Signatures (LINCS) [30], and Rapid Acceleration of Diagnostics (RADx) [32]. Moreover, CEDAR is also being deployed at 88 healthcare sites in 8 African countries as part of the Virus Outbreak Data Network (VODAN) [33, 34]. In all of these activities, the primary goal is to drive the authoring of "rich," standards-adherent, machine-actionable metadata to ensure that the associated datasets will be FAIR.

Evaluating Metadata with Templates: The FAIRware Workbench
Most scientific metadata are crafted in an informal manner, and the vast majority of datasets, consequently, are not FAIR. Metadata authors typically do not adhere to community standards, and they tend to provide only a minimal number of metadata [35]. Thus, third parties, including funders and publishers, often want to assess how FAIR datasets actually are. When metadata do not adhere to community standards, it would be helpful to know if there are particular standards that are simply hard for researchers to use reliably. Furthermore, many investigators would like to know how to make their metadata more standards-adherent, taking advantage of reporting guidelines and ontologies that they may have overlooked. The FAIRware Workbench is a system currently in prototype form that was created with all these use cases in mind.

The FAIRware Workbench takes as input (1) pointers to a set of online dataset records to be evaluated (generally a list of DOIs) and, optionally, (2) a metadata template (taken from the CEDAR template library). It generates as output (1) an analysis of how well the metadata in the datasets adhere to the standards encoded in the metadata template, indicating which template fields are most often noncompliant and (2) new, candidate versions of the metadata records that demonstrate better adherence to the standards represented by the template. The system thus evaluates in batch a set of datasets from a target repository, checks the metadata for standards adherence, reports on situations where standards were used inappropriately or omitted entirely, and produces, for each problematic metadata record that it detects in the original set, an alternative record that the system suggests adheres better to the appropriate standards. When an appropriate CEDAR metadata template is available for reference, because it will have been created to reflect the standards of the relevant scientific community, the template serves as the "gold standard" against which to benchmark the input metadata in the FAIRware-based evaluation.

**Metadata Evaluation Report**

Found **3** issues

Template: Sample Section

| FIELD NAME | FIELD VALUE | ISSUE | SUGGESTED REPAIR |
|---|---|---|---|
| preparation_medium | "" | MISSING_REQUIRED_VALUE | |
| source_storage_time_value | "208 days" | EXPECTING_INPUT_NUMBER | 208 |
| storage_medium | "Cryostat embedded" | VALUE_NOT_ONTOLOGY_TERM | OCT Embedded |

Dropdown options: OCT Embedded, Buffered Formalin (10% NBF), PFA 4%, 1 x PBS, CMC Embedded, OCT Embedded Cryoprotected (sucrose), Paraffin Embedded

**Figure 3:** FAIRware Workbench analysis of a metadata record for a tissue sample. The screen capture shows the analysis of one of the records in the repository, indicating where the reporting guideline may not have been followed or where ontology terms were not used appropriately. The system automatically corrects the string "208 days" to the integer 208. There is no obvious correction for the entry for "storage medium." Because in this example the FAIRware workbench is in interactive mode, it offers the user a menu of ontology terms that might provide a standards-adherent value.

Figure 3 shows an analysis that the system has performed on a collection of datasets from the HuBMAP project—in this case indicating potential errors in one of the metadata records describing the section of a biological sample from which a specimen was taken. By comparing each metadata record in the repository against the standard metadata template, the FAIRware system acts like a "spell checker" or debugging tool to offer suggestions on how to improve the metadata [36]. When the template specifies that a metadata field is to be filled with a term from a standard ontology, the software verifies that the value for that field is indeed a term from that ontology. If the value does not match an ontology term, then the FAIRware system identifies the *closest match* in the ontology that it can find and it assumes that the metadata author intended to use the identified term, suggesting that the user replace the aberrant value with the term from the ontology. The software also verifies that the required *field names* enumerated in the template are present in the metadata record under evaluation. If a required field appears to be missing, then the FAIRware system will determine whether a field name in the current record appears to be a misspelling of the absent field name. If so, then the software will suggest correcting the assumed misspelling. Users have the option of requesting that FAIRware always implement its suggestions for metadata improvements automatically, or of having it do so only with explicit user approval on a record-by-record basis.

FAIRware is able to make certain judgments about legacy metadata even if the user is unable to indicate a candidate template for reference. For example, it may be possible to find in the CEDAR library a metadata template whose fields bear similarity to many of those listed in the metadata under evaluation. In that case, the system can attempt to use the template as a candidate reference for the metadata under evaluation. Even when no reference template is available, FAIRware can assess how well strings in the metadata under evaluation match terms in known ontologies. The system then can make guesses regarding the possibility that the original metadata author may have intended to use a standard term and inadvertently failed to do so.

As the FAIRware Workbench evaluates the metadata records that are input into the system, it stores in an output database the "cleaned up" metadata based on its analysis of the original records and the gold-standard metadata template. In principle, the metadata in the more standards-adherent output database should be searchable with greater precision and recall than is possible with the original records. As a result, using the "cleaned up" records when searching for datasets could provide a means to make the original datasets functionally more FAIR. We never replace the original records with our new metadata because, like a traditional spell-checker, FAIRware can inadvertently introduce new errors in its attempt to correct what it identifies as mistakes. Instead, we make the FAIRware version of the metadata available as a separate resource for third parties who wish to perform dataset search and to ensure that their queries have maximum recall.

The FAIRware Workbench generates summary information regarding the input metadata as a whole (Figure 4). The system creates reports that enable users to visualize the degree to which the metadata under evaluation include the fields suggested by the CEDAR template, the degree to which metadata values are taken from appropriate ontologies, and which metadata fields are least likely to be filled with an ontology term. Users with oversight responsibilities for data stewardship thus can obtain a perception of overall metadata quality in the input datasets, and they can learn which reporting guidelines (and which fields of reporting guidelines) may be difficult for investigators to apply. Such results can inform the evolution and improvement of the standards adopted by scientific communities.

**Methods**
Technology such as CEDAR and FAIRware, which both rely on the ability to represent metadata in terms of reusable templates, motivates our work to develop methods for capturing metadata standards as templates and for encoding them in a machine-actionable manner. These methods allow us to anticipate an ecosystem for open science that emphasizes standard templates as the basis for disseminating metadata standards and for ascertaining and ensuring data FAIRness.

Formulating Metadata: Metadata for Machines Workshops
FAIRware and CEDAR both operate on the same machine-readable representations of metadata templates. The experience with these two systems demonstrates the value of having metadata standards available in a format that can "plug and play" with alternative software applications. The question, of course, is how does one know what to include in a metadata template in the first place?

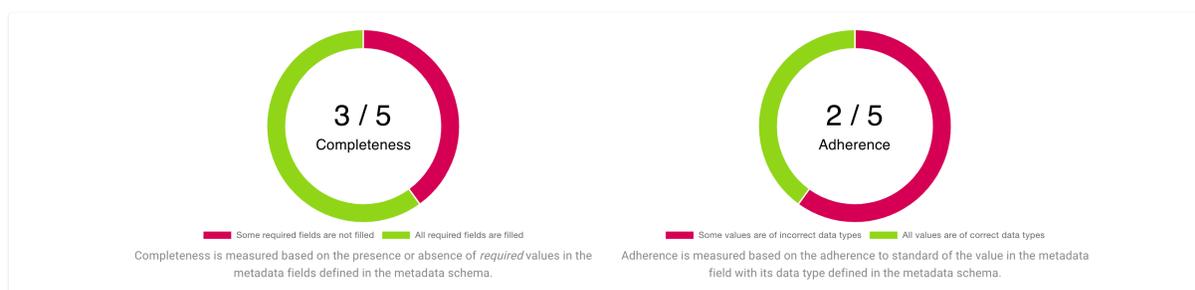

**Figure 4:** FAIRware Workbench summary analysis. The Workbench provides the user with an overview of how well the input data adhere to the standard defined by the metadata template indicated at runtime. We can see that, overall, there are many records with missing required fields, and several records with field values that do not adhere to standards (such as the use of standard ontology terms). At the bottom of the screen, users can see more detail and review which metadata fields cause the most difficulty.

The situation may not require an inordinate amount of work when an appropriate reporting guideline already exists, such as MIAME or MIATA. The problem, then, is to transform the elements of each checklist into discrete template attributes (field names), and to assign a value type to each field. When the value type is one of "character string," the developer needs to determine whether the value should be selected from a pre-enumerated list and, if so, whether there is an existing ontology or common data element that should be used.

Sometimes, the template does not reflect a standard reporting guideline for a particular class of experiment but, instead, reflects the built-in metadata fields that are anticipated by a particular repository. For example, domain-specific databases at the U.S. National Center for Biotechnology Information, such as BioSample and BioProject, set up explicit expectations for their metadata fields [37], and it often makes sense to create metadata templates that provide the list of metadata fields that the particular repositories require.

Often, however, there is no existing reporting guideline for the area of study, and there is no repository that can provide target metadata fields for the necessary template. In that case, developers need to create a new template—either from scratch, or by modifying elements of existing templates. Creating such a template becomes an engineering problem, where the goal is not only to identify the essential elements of the class of scientific experiments that need to be described (the work that would be associated with authoring a reporting guideline in general), but also to represent those elements in a form that can be encoded using the formal structure assumed by our two applications.

The GO FAIR Foundation (Leiden, the Netherlands) has formalized the process of engineering new metadata templates, with the explicit purpose of scaling the creation of domain-relevant, standards-based metadata. GO FAIR has created the idea of "Metadata for Machines Workshops" (M4Ms) to help investigators to create computer-stored templates that are compatible with both the CEDAR and FAIRware workbenches. M4Ms bring together stakeholders in particular areas of science to craft machine-actionable metadata templates that are fit for purpose and, if necessary, ontologies that supply terms in support of those templates. These workshops are remarkably efficient—typically lasting one to three days—and they produce rich metadata specifications that, more importantly, can evolve over time. Although some well-known reporting guidelines such as MIAME took months or years to work out, M4Ms are extremely compressed events, almost like "hackathons," that owe their productivity to multiple factors. First, the participants' familiarity with previously designed scientific reporting guidelines prepares them for the work before them. Second, participants come primed to work intensively, and the collaborative, interactive environment created by the M4M facilitators helps to tease out problems and to focus attention on issues that require consensus-building. Third, the workshop incorporates highly usable metadata and vocabulary modeling technologies such as the CEDAR Workbench to provide a shared, interactive platform where the workshop participants can prototype metadata records and accompanying ontologies, and where they can observe the ramifications of their proposals. Finally, building a structured, computer-based reporting guideline has similarities to the crafting of any declarative, machine-readable knowledge base for an intelligent system; the knowledge-elicitation techniques that were pioneered by the artificial intelligence community in the 1980s and 1990s turn out to be helpful in the context of M4M workshops [38, 39].

The GO FAIR initiative has now hosted dozens of M4M workshops, in domains as varied as studies of antimicrobial resistance, neuroimaging for experiments related to human consciousness, and evaluation of technology to capture wind energy [40]. The M4M facilitators have also demonstrated that their format can be replicated by other facilitators who are trained in the approach. GO FAIR construes M4M workshops as the vehicle for creating key elements of what it calls *FAIR Implementation Profiles* (FIPs) [41]—an intentionally generic term for the set of design choices that a given scientific community makes to implement the FAIR Guiding Principles. These design choices largely boil down to decisions about standards. From a practical perspective, the design choices inherent in a metadata template provide the FIP elements over which a group of investigators has direct control: the relevant reporting guideline(s), any additional metadata elements, and the ontologies used to provide controlled terms. Hence, these elements, which are documented in the content stored in CEDAR and BioPortal, can be referenced as FAIR-enabling resources that contribute to the overall FIP. The processes of M4M workshops provide the foundation needed to translate the abstract FAIR Guiding Principles practically and efficiently into concrete, computer-stored solutions that can ensure data FAIRness.

Of course, metadata templates do not need to be developed in the setting of formal workshops. Many scientific consortia have been very successful creating unambiguous metadata specifications without the need for external facilitators or structured processes. The Adaptive Immune Receptor Repertoire (AIRR) community [31] and the HuBMAP consortium mentioned previously [22] are examples of two research organizations that have been successful in creating CEDAR-compliant metadata standards in a more organic fashion.

The key observation is that CEDAR metadata templates necessarily adhere to the requirements and preferences of the scientific communities that create them. Fields are included in a template because the communities view them as essential. Preferences regarding the distinctions to be made (1) about classes of experiments, (2) about the ontologies to use to specify those distinctions, and (3) about the granularity with which data need to be described are built into the design of the template. The templates necessarily reflect the values and beliefs of the given communities of practice that built the templates in the first place, and the templates provide a vehicle for communication—both within and without the community—that offers more precision and more perspicuity than is possible with reporting guidelines that are captured only in prose.

Encoding Metadata: A Machine-Actionable Model

CEDAR and FAIRware are able to operate with the same metadata descriptions because the metadata are encoded in a shared, machine-readable format. Our approach starts with a lightweight, abstract model of the core characteristics of metadata. The model provides a consistent, interoperable framework for defining metadata templates and for creating and filling in instances of metadata that comport with those templates [42].

The model assumes that templates are assembled from more granular *template elements*, which are themselves templates. Template elements are thus building blocks of more elaborate templates. The template elements contain one or more pieces of information, such as a text field or a date field, or a subordinate template element. The fields in a template represent atomic pieces of metadata, and they are rendered as blanks to be filled in by the CEDAR metadata editor (as in Figure 1). The model's nested approach to defining template elements allows fields such as *phone number* and *email* to be contained in a template element called Contact Information, which could itself be contained in a template element called Person. The model also indicates possible value types for metadata fields, as well as the designation that fields have textual descriptions and, in keeping with the FAIR principles, unique identifiers.

Our approach uses JavaScript Object Notation (JSON), an open, standard file format widely used for data interchange. JSON is reasonably human-readable, and it is well suited for encoding the attribute–value pairs that characterize scientific metadata. Our model of metadata templates, template elements, and template fields is represented in a JSON-based specification known as *JSON Schema*, which provides a structural specification of the model's components. Metadata *instances* (i.e., the metadata for particular experiments) are encoded using JSON-LD, in accordance with the overall template model represented in JSON Schema. JSON-LD is a JSON-based format that adds support for references to "linked data" (i.e., persistent identifiers on the Web), which in our case provide semantic descriptions of metadata entries through the use of the persistent identifiers associated with standard ontology terms. Thus, in the JSON-LD

```
    "sample_ID": {
      "@value": "Visium_9OLC_I4_S2"
    },
    "source_storage_time_value": {
      "@value": "208",
      "@type": "xsd:float"
    },
    "preparation_medium": {
      "@id": "http://purl.bioontology.org/ontology/MESH/D000432",
      "rdfs:label": "Methanol"
    },
    "processing_time_value": {
      "@value": "4",
      "@type": "xsd:float"
    },
    "processing_time_unit": {
      "@id": "http://purl.obolibrary.org/obo/UO_0000031",
      "rdfs:label": "minute"
    },
    "storage_medium": {
      "@id": "https://purl.org/hubmapvoc/samples-voc-additions/OCTEmbedded",
      "rdfs:label": "OCT Embedded"
    },
    "storage_temperature": {
      "@id": "http://ncicb.nci.nih.gov/xml/owl/EVS/Thesaurus.owl#C185336",
      "rdfs:label": "-80 Degrees Celsius"
    }
}
```

**Figure 5:** The JSON-LD representation of the HuBMAP metadata seen in Figure 1. The explicit incorporation of the persistent identifiers of ontology terms provides a semantic foundation for the corresponding metadata fields. We can see, for example, that the value for "preparation medium" refers to a term from MeSH.

representation in Figure 5, the value for *preparation_medium* has a precise, resolvable meaning because, in accordance with the metadata template, it relates to a specific term in the Medical Subject Headings (MeSH) vocabulary. The value for *processing_time_unit* gets its semantics from the OBO Units Ontology. The value for *storage_temperature* gets its meaning from the National Cancer Institute Thesaurus. When the JSON-LD representation indicates that the *preparation_medium* for the sample is *methanol* as defined by MeSH, then a computer can reason about the metadata and about the MeSH ontology (and about other ontologies mapped to MeSH) to conclude that the methanol preparation medium is an *alcohol*, or even that the preparation medium is highly flammable and also highly toxic. The approach enables logical analysis of the metadata and of the experiment that the metadata describe in ways that are not possible with simpler file formats designed primarily for human users.

Although the JSON-LD encoding for metadata can appear a bit daunting, typical users never see the machine-readable representation. Scientists create metadata using CEDAR Web forms that are simple and intuitive, and they review the metadata in legacy datasets using FAIRware in a manner that shields them from the internal format. Nevertheless, because the JSON-LD is a standard data exchange format that is readily parsed by computers, it provides a straightforward, lightweight mechanism to encode metadata with clear semantics and to distribute the metadata to a variety of applications.

**Discussion**
We have developed a model for scientific metadata and we have made that model usable by both CEDAR and FAIRware. Our approach shows that a formal metadata model can standardize reporting guidelines and that it can enable separate software systems to assist (1) in the authoring of standards-adherent metadata and (2) in the evaluation of existing metadata. By creating a framework that supports scientific communities directly in their work to codify which reporting guidelines and which ontologies to use in their data annotation, we finesse the impossible task of determining, out of context, whether metadata are "rich," whether they adhere to the proper standards, and whether they use ontologies that are themselves FAIR. Instead, we adopt a formal computer-stored representation that enables different communities of practice to build domain-specific templates for the types of metadata that are appropriate for serving the needs of their disciplines.

Formal Knowledge Bases That Capture Community Standards
In a series of invitational workshops, Gregory and Hodson [43] have proposed a Cross Disciplinary Interoperability Framework (CDIF) as a *lingua franca* for creating metadata to support FAIR data reuse. Unlike the work presented in this paper, CDIF focusses on generic, domain-independent implementation standards, offering a set of guidelines and best practices for representing metadata at a level of specificity that is helpful to software developers. CDIF proposes general-purpose standards that might be used to encode metadata for purposes of accessibility (e.g., the use of JSON, in the case of CEDAR templates), but the framework makes no commitment to the metadata content needed for investigators to find datasets using discipline-specific terms and to understand the data in a manner that allows for reuse and for interoperability with other research results. As work on CDIF progresses, however, it may be appropriate for the implementation of metadata templates to evolve to incorporate ideas that originate from this effort.

CDIF is a component of broad work in the FAIR community to define FAIR Digital Objects (FDOs)—machine-interpretable encapsulations of data, metadata, analytic methods, and other products of research—that might integrate into larger software systems. FDOs aim to address the overall goals of managing and using FAIR data, but FDOs are software artifacts that are at a level of abstraction beneath that of the community standards that are the focus of the metadata templates discussed in this paper. CEDAR templates certainly need to operate in a software environment where developers can make commitments to the necessary implementation standards, but such standards are quite distinct from community-based metadata standards that capture the *knowledge* of the minimal set of descriptors needed to make sense of a particular type of experiment and the standard terms from which investigators might fashion those descriptions. This distinction between knowledge and implementation is one with which the artificial intelligence community has struggled for decades [44].

Our template model provides a straightforward mechanism to translate the knowledge of textual reporting guidelines into a machine-actionable format. Because the model can be easily read and processed by a variety of applications (mitigating vendor lock-in and associated blocks to interoperability), it can form the basis for a standard, technology-independent means to communicate reporting guidelines in a computable fashion. Our intention is not to introduce yet another redundant data standard to make an already complex landscape of standards even more confusing [45]. Rather, we aim to provide a mechanism for rendering an existing type of standard (namely, reporting guidelines) more precise and more readily actionable. There is no agreed-upon convention for how reporting guidelines should be rendered, and the availability of a coherent format that is compatible with widely used knowledge-representation standards provides an obvious advantage.

This approach enables groups of investigators to build into their templates the practice-centered elements of data FAIRness. Existing automated approaches that attempt to assess whether data are FAIR break down due to their inability to determine algorithmically the tacit data-annotation practices of particular research communities. However, when investigators construct appropriate templates, CEDAR supports the authoring of FAIR data and FAIRware supports the assessment of data FAIRness because the data-annotation practices of the relevant communities are encoded directly into the metadata templates. The standards endorsed by the community are precisely those standards that are used to produce the template. The fields selected for the template are precisely those fields that the community believes are required for the metadata to be adequately "rich." Inevitable revisions to community metadata standards can be addressed simply by editing existing CEDAR templates.

The templates in our work can each be viewed as a knowledge base of the preferred data-annotation practices of a particular scientific community for a particular class of scientific experiment. In that sense, our methods build on approaches that have been practiced for decades to build knowledge-based systems [46]. The M4M workshops used to create metadata templates adopt techniques that are reminiscent of the knowledge-elicitation methods studied by the knowledge-engineering community [38, 39]. These workers have built electronic knowledge bases to "plug and play" with different problem solvers [47], enabling constraint-satisfaction engines or classification systems or planners to reason about different aspects of the same knowledge base. By analogy, our metadata templates should be able to interoperate with a variety of applications well beyond CEDAR and FAIRware. One can speculate about computer programs that could use metadata templates to translate metadata from one community standard to another automatically, or to reason about how datasets with disparate annotations might be harmonized.

Our experience with CEDAR and FAIRware demonstrates the value of using formal knowledge-representation methods to characterize information about the annotation of experimental datasets. One would not necessarily predict the effectiveness of this approach from the history of the well-known MIAME standard, however. The Microarray and Gene Expression Data Society (MGED), after promoting the wildly successful MIAME minimal information guideline, went on to create the formal MicroArray Gene Expression Object Model (MAGE-OM) in UML [48], and the even more formal MGED Ontology in OWL [49]. Neither of these resources gained much traction in the microarray community. They were complicated and beyond the understanding of most practitioners, and there were no easy-to-use tools that could insulate investigators from that complexity—until MAGE-OM was later folded into a spreadsheet-based application known as MAGE-Tab [50]. Although CEDAR metadata are encoded in JSON-LD, our applications do not expose users to JSON-LD syntax, and users generally view the templates only through intuitive user interfaces. Our experience emphasizes the long-established importance of coupling formal, coded knowledge-representation approaches with applications that can shield typical users from the intricacies of the underlying models.

Deciding What to Model
Minimal information checklists such as MIAME have been criticized for their lack of specificity and inadequate granularity [51]. CEDAR metadata templates, on the other hand, can be created with explicit fields that can refer to arbitrarily fine-grained properties of an experiment. The metadata templates impose a rigidity on the description of experiments that may make some data curators feel uncomfortably constrained, and the templates may ask for details that the curators may not be able to provide. The challenge is to create templates that make minimal assumptions about the metadata that curators will be able to specify, but that can anticipate (1) the attributes of datasets for which third parties may want to query, (2) the features of the experimental situation that need to be described to ensure replicability, and (3) the attributes of datasets that need to be indicated to ensure appropriate interoperability with other datasets. Scientists will naturally prefer to specify the most minimal metadata that they can in the most flexible manner possible. A significant risk of our approach is that template developers will be overambitious in their beliefs regarding what is "minimal" information, leading to metadata models that make too many expectations of investigators sharing their data. In practice, CEDAR has not been well received when scientists have been asked to provide what they consider to be overwhelming amounts of information about their experiments. On the other hand, for data to be FAIR, it is important for datasets to be searchable in domain-specific terms and to describe experimental conditions adequately. Researchers have learned that publishing a research paper requires them to adhere to certain conventions, and they are beginning to accept that "publishing" their datasets in a FAIR manner also requires adherence to certain professional norms.

Our approach bears similarities to that of the ISA software suite, which models Investigations, Studies, and Assays [52]. The ISA suite of tools enables investigators to create metadata stored as separate files that reflect the investigations (overall projects), studies (experiments), and assays (measurements performed as part of an experiment) associated with their research. The ISA approach provides a general model of a kind of biological investigation, and it enables curators to create metadata that relate to the different aspects of the overall model. ISA models a specific class of scientific *investigation*, however, unlike our model, which models *metadata*.

The model of metadata facilitates the creation of detailed metadata templates that explicitly specify the datatype of each metadata field and, when appropriate, the specific ontologies (or parts of ontologies) that should be used to provide values for that field. The CEDAR approach enables our metadata-authoring system to generate user-friendly Web forms directly from a metadata template without the need for any computer programming, and it creates precise JSON-LD from the user's entries. CEDAR does not hard code into its template language distinctions about the components of scientific investigations such as those represented in the ISA model. Instead, our template model emphasizes only what needs to be said about metadata—and thus it is applicable to fashioning metadata for all kinds of FAIR digital research objects (datasets and beyond). When users wish to model the structure of research projects in CEDAR, they may do so by creating template elements that correspond to different aspects of the entities modeled by ISA. Thus, the structure built into ISA can be clearly expressed through design choices that CEDAR template developers make when creating template subunits, but the developer is not forced to adhere to ISA, or to any other overarching framework.

Because CEDAR templates define models of metadata, the templates can structure annotations for experiments that do not comply with the investigation-study-assay framework. For example, CEDAR's generic template model has accommodated descriptions of observational studies related to wind energy in Denmark [53] and HIV infections in Africa [33, 34]. The HuBMAP metadata that we present in this paper, which assume a donor–organ–specimen–assay organization, has no obvious means to fit into an ISA model.

If the goal is simply to facilitate enhanced annotation of a singular, very specific kind of study data, then one does not need the sophistication of the technology that CEDAR offers. For example, developers associated with the NIH Biomedical Informatics Research Network (BIRN) created an attractive, hardcoded, interface for annotating their particular kinds of neuroimaging data as part of their Digital Asset Management System (DAMS) [54]. The Stanford Microarray Database developed a similarly well-crafted tool, known as Annotare [55], for entering metadata about gene-expression datasets in accordance with hardcoded MAGE-Tab descriptions [50]. Such bespoke systems have been served their users very well through interfaces that reflect the details of their particular experimental domains, but these systems have little to offer users beyond those for whom the systems were first built and whose assumptions drive the performance of the tools.

By taking on completely a *model-independent* view of the task of data annotation, Google has developed a general-purpose search engine for online datasets [56] that enables retrieval of arbitrary Web pages where datasets are highlighted with keywords from the Schema.org terminology [57]. Schema.org is a comprehensive, shared terminology first proposed for aiding search across Google, Bing, Yahoo!, and Yandex to perform online queries and to facilitate the presentation to users of relevant results. Rather than trying to sort through multiple domain-specific ontologies, Google's Dataset Search engine limits itself to finding datasets for which the experimenters have used Schema.org as the basis for adding annotation. Although this facility appears to be very useful, such Schema.org representations are not necessarily endorsed as standards by any scientific community, and the annotations rarely are "rich" with respect to any particular research domain. The biology community is working to extend Schema.org through the Bioschemas initiative [58], but activity to enhance Schema.org for use within other areas of science is rare. In the end, the use of Schema.org in online metadata will make those metadata searchable using Google, but it will not make the corresponding datasets FAIR.

Thus, the landscape of automated approaches to promote data stewardship is dotted with standard software objects that encapsulate relevant information (as with CDIF [43]); the use of narrow, domain-specific ontologies within highly customized user interfaces (as with Annotare [55]); the use of broad, Web-encompassing ontologies (as with Google Dataset Search [56]); and the use of models of certain classes of interventional experiments (as with ISA [52]). Consequently, the landscape is replete with nonintegrated tools that attempt to address many different aspects of creating metadata and other digital research objects. But not one of these tools, on its own, can ensure that metadata adhere broadly to arbitrary community standards, that metadata are rich, and that the metadata use terms from ontologies that are themselves FAIR.

A Pathway to FAIR Data
The FAIR principles have provided a valuable way to think about the qualities of data that make the data sharable and useful to other investigators. As we have emphasized, the challenge is that many of the criteria cannot be evaluated by simply examining online datasets, since the FAIR principles depend inherently on subjective criteria. Because the whole idea of FAIRness is tied to the particular beliefs of disparate scientific communities, the only way to evaluate online datasets for FAIRness computationally is to codify those beliefs in some machine-actionable form.

In our approach, we view metadata templates and the ontology terms used to populate those templates as the vehicles by which we capture community standards and communicate them throughout our ecosystem. Because the metadata templates can encode all relevant FAIR-related standards, the templates can transmit those standards directly from the relevant communities and deliver them to software systems such as CEDAR and FAIRware that act on those standards.

As developers, we do not take a position on which distinctions about metadata a community needs to include in a particular template; the metadata content solely reflects the discretion of the community members. Similarly, it is up to the community to determine what makes metadata "rich." There is no second-guessing the preferences of the given scientific community; everything is recorded directly in the template.

System builders who have attempted to develop automated systems to assess the FAIRness of online datasets have been hindered by the problem that so many of the FAIR Guiding Principles are dependent on subjective criteria, which often seem elusive and unprincipled to observers outside the relevant scientific communities. The discrepancy in the performance of different computer-based FAIR evaluation tools when applied to the same datasets [60] may be a consequence of the diverse ways in which the tools attempt to respond to these subjective criteria. Some FAIR evaluation systems incorporate the subjective beliefs of particular communities into specialized assessment rules. For example, FAIRshake [5] enables users to plug in discipline-specific *rubrics* that codify the community-specific constraints needed to assess particular digital objects. Users of the tool need to identify the relevant communities according to whose beliefs the evaluation should take place, and to identify (or create) a rubric that reflects those beliefs. Such rubrics are useful only in conjunction with the evaluation tool, and users of the tool need to have sufficient insight to know when a new rubric might be required. This situation places a significant responsibility on the evaluation-tool user if the tool is to generate credible results.

The CEDAR approach, on the other hand, shifts the responsibility to the scientific community, which obtains a direct benefit from creating the required metadata templates in the first place. The templates can be used by the FAIRware Workbench as a mechanism to evaluate data FAIRness, and they also can be used by CEDAR to help author new metadata that are guaranteed to make the corresponding datasets FAIR. The FAIRware Workbench offers additional benefits to the scientific community, such as identifying which community standards appear to be hard to use and offering specific suggestions for how to make the metadata annotating legacy datasets more adherent to the identified standards.

As envisioned by the FAIR principles, the ability to encode community standards in an examinable, semantically rigorous, machine-actionable format creates opportunities throughout the research enterprise. Metadata templates, such as those used by CEDAR and FAIRware, can provide canonical representations of the reporting guidelines (and associated ontologies) important to a given community, and they can enable both people and computers to access those representations within an ecosystem of tools that can enhance data FAIRness. Such templates enable rigorous specification of metadata in advance of dataset production, and they support evaluation and correction of existing metadata with respect to community-provided standards after datasets are archived.

We do not discount the considerable effort required to create such templates in the first place. To satisfy funding and regulatory mandates for FAIR data, scientific communities will need to mobilize to create the necessary standards and to commit to the application of those standards as part of routine data stewardship [61]. The inherent rigor, precision, and reusability that accrues from machine-actionable metadata templates can support this activity, with the goal of leading to better data and, ultimately, to better science.

**Data Availability**
HuBMAP datasets are accessible through the HuBMAP Portal (see https://portal.hubmapconsortium.org). The HuBMAP metadata that we have processed using the FAIRware workbench can be reviewed at https://fairware.metadatacenter.org/publish-data/index.html.

**Code Availability**

The software described in this paper is available for execution at http://metadatacenter.org. All source code is available from the Metadata Center landing page at https://github.com/metadatacenter.

**Acknowledgments**

This work was supported by a contract on behalf of the Research on Research Institute, with funding from the Wellcome Trust, the Austrian Science Fund, the Canadian Institutes of Health Research, the National Institute of Health and Care Research (UK), and the Swiss National Science Foundation; by grant R01 LM013498 from the U.S. National Library of Medicine; by grant U24 GM143402 from the U.S. National Institute of General Medical Sciences; by award OT2 OD033759 from the U.S. National Institutes of Health (NIH) Common Fund; and by award OT2 DB000009 from the NIH Office of Data Science Strategy. We are grateful to Stephen Fisher for providing us with legacy HuBMAP datasets, to Sunteasja Billings for creating the initial templates for HuBMAP metadata, and to Michelle Barker and Peter Kant for helpful discussions.

**Author Contributions**

All authors participated in the work. MAM wrote the initial draft of the manuscript. All authors reviewed and edited the manuscript.

**Competing Interests**

ES is the Scientific Director of Partners in FAIR (https://partnersinfair.com). There are no other competing interests.